\RequirePackage[T1]{fontenc}
\documentclass[12pt]{article}

\usepackage[height=8.85in,width=6.45in]{geometry}

\usepackage{times}

\usepackage[utf8]{inputenc}
\usepackage{amsmath}
\usepackage{amssymb}
\usepackage{mathtools}
\numberwithin{equation}{section}
\usepackage{slashed}
\usepackage{braket}
\usepackage[svgnames]{xcolor}
\usepackage[colorlinks,citecolor=DarkGreen,linkcolor=FireBrick,linktocpage,pagebackref]{hyperref}
\usepackage{url}
\usepackage{cite}
\usepackage{graphicx}
\usepackage{tikz}
\usepackage{tikz-cd}
\usepackage{spectralsequences}

\usepackage{courier}
\usepackage{bm}
\usepackage{dashbox}
\usepackage{subcaption}
\usepackage{footmisc}

\usepackage{mdframed}

\usepackage{blkarray}
\usepackage{arydshln}
\usepackage{dsfont}
\usetikzlibrary{patterns}

\renewcommand{\title}[1]{\vbox{\center\LARGE{#1}}\vspace{5mm}}
\renewcommand{\author}[1]{\vbox{\center#1}\vspace{5mm}}
\newcommand{\address}[1]{\vbox{\center\em#1}}
\newcommand{\email}[1]{\vbox{\center\tt#1}\vspace{5mm}}

\begin{document}

\begin{titlepage}

\vspace*{0.5cm}

\title{Higher Connection in Open String Field Theory}

\author{Yichul Choi}

        \address{School of Natural Sciences, Institute for Advanced Study, Princeton, NJ, USA}
        \email{choi@ias.edu}

\abstract

We define a 2-form connection in the space of classical solutions of the bosonic open string field theory, using the open string star product and integration. The corresponding higher holonomies and the 3-form curvature are new observables invariant under the infinite-dimensional gauge algebra of open string field theory. The definition is analogous to that of Berry phase in quantum mechanics and is motivated by recent studies on higher Berry phase in condensed matter physics and quantum field theory. We suggest identifying this 2-form connection with the Kalb-Ramond $B$-field of the closed string background at least in favorable situations. Also discussed are sigma models whose target space is the moduli space of conformal boundary conditions of a two-dimensional CFT with the $B$-field given by a cousin of this 2-form connection.

\end{titlepage}

\tableofcontents

\section{Introduction}

What is the minimum set of data of a conformal field theory (CFT) from which all possible correlation functions on arbitrary manifolds can in principle be deduced?
It is commonly said that the spectrum of local operators and their operator product expansion (OPE) coefficients, consistent with the conformal bootstrap axiom (crossing symmetry of 4-point correlation functions), define a CFT.
However, this cannot be the full answer for at least two reasons. 
First, there exist distinct CFTs which share exactly the same set of local operators and OPE coefficients, and yet differ by the spectrum of extended operators and defects.\footnote{Examples are the four-dimensional maximally supersymmetric Yang-Mills theories with different global structures of the gauge group which all share the same Lie algebra \cite{Aharony:2013hda}.}
Second, a solution to the bootstrap axiom does not necessarily give enough information to define the CFT on manifolds which may have a complicated nontrivial topology.\footnote{The local operator data allow one to compute, for instance, correlation functions on a Euclidean sphere $S^d$ and the partition function on $S^{d-1} \times S^1$. But they are in general not enough to determine the partition function of the theory, for instance, on $S^{d-2} \times S^2$. In two-dimensions, a solution to the conformal bootstrap axiom on a four-punctured sphere does not necessarily define a CFT consistent on higher genus surfaces, and one needs to impose additional constraints such as modular invariance.}
In general spacetime dimensions, the answer to the question is yet to be known, and there may not be a unique answer.

In two-dimensional CFTs, the question bears potentially much deeper physical significance, and a very interesting answer is known at least in the case of rational conformal field theories (RCFTs).
It potentially has deeper physical significance because of the idea that, when viewed as a part of the worldsheet CFT in perturbative string theory, two-dimensional CFTs may be regarded as representing a string-theoretic generalization of the notion of spacetime geometry.
More precisely, they would represent a class of ``on-shell'' geometries for the strings.
Finding a minimum set of data from which the full two-dimensional CFT can be recovered is therefore connected to the question of understanding the precise mathematical formulation of stringy geometry. 

In the case of RCFTs, thanks to the series of works by Fjelstad, Fuchs, Runkel and Schweigert \cite{Fuchs:2002cm,Fuchs:2003id,Fuchs:2004dz,Fuchs:2004xi,Fjelstad:2005ua,Fjelstad:2006aw}, one arrives at the following surprising fact:
The full data of a rational conformal field theory is determined by the choice of a chiral algebra and a consistent solution to the \emph{boundary} bootstrap axiom, that is, the crossing symmetry of the four-point disk correlation functions of boundary local operators without any bulk operator insertions.
Knowing the spectrum of boundary primary operators on some fixed conformal boundary condition and their OPE coefficients is then enough to derive the spectrum of all bulk primary operators, as well as all the correlation functions on arbitrary higher genus surfaces.
The proof of the statement utilizes the theory of modular tensor category (MTC) and the corresponding three-dimensional topological quantum field theories (TQFTs) \cite{Moore:1988qv,Witten:1988hf,Elitzur:1989nr}, and it will not be reviewed here.\footnote{
Let us briefly outline the intuition behind this for the experts.
A solution to the boundary bootstrap axiom on a rational conformal boundary gives rise to a generally non-commutative ``algebra object'' internal to the MTC of the chosen chiral algebra. With a few assumptions, this algebra object can then be used to define a topological surface defect in the 3d TQFT characterized by the MTC \cite{Kapustin:2010if,Roumpedakis:2022aik}. The RCFT is obtained by the dimensional reduction of the TQFT on an interval, where on the two boundaries of the interval live the holomorphic and anti-holomorphic operators, respectively, and at the middle of the interval one inserts the topological surface defect.
Different rational conformal boundary conditions of a given RCFT give rise to ``Morita-equivalent'' algebra objects of the MTC, and they all lead to the same surface defect of the TQFT.

Another, more common way to characterize an RCFT is by the choice of a chiral algebra and a consistent spectrum of bulk primary operators satisfying the bootstrap axiom on a four-punctured sphere and the modular covariance of torus one-point functions \cite{Moore:1988qv}. We may regard these different ways of characterizing an RCFT as a version of open/closed duality. 
}

Our eventual aspiration is to pursue a similar structure in more general (two-dimensional) CFTs which may not be rational, or even in rational theories, to find different ways in which somehow the data of boundary conditions and boundary correlation functions are enough to fully reconstruct the entire CFT.
The results from a recent work \cite{Choi:2025ebk} suggest that in some familiar non-linear sigma models, the target space geometry, not just the metric but also the Kalb-Ramond $B$-field, is fully encoded in certain boundary correlation functions.
In particular, one can identify a subset of the space of conformal boundary conditions as the target space of the theory, with the
metric given by the Zamolodchikov metric \cite{Zamolodchikov:1986gt,Seiberg:1988pf,Kutasov:1988xb} and the $B$-field extracted from 3-point correlation functions of boundary-condition-changing operators \cite{Choi:2025ebk}.
This will be explained later.
In these theories, therefore, there is again a sense in which the full data of the CFT is encoded in the information about boundary conditions and boundary correlation functions.

An elegant perspective is provided by Witten's cubic open string field theory \cite{Witten:1985cc}. 
In open string field theory, one takes as an input a consistent worldsheet CFT with a choice of a conformal boundary condition.\footnote{Progress in the past has shown that the open string field theory contains non-perturbative information about the theory \cite{Schnabl:2005gv,Erler:2019vhl}.}
The worldsheet boundary correlation functions are believed to organize into a non-commutative algebra, sometimes called the ``open string star algebra.'' 
Other conformal boundary conditions of the worldsheet CFT appear as classical solutions to the equation of motion \cite{Erler:2014eqa,Kiermaier:2010cf,Erler:2019fye,Erler:2019vhl} and open string field theories formulated with different choices of the input conformal boundary condition of a given worldsheet CFT are all equivalent.
Therefore, the non-commutative algebra is really a characteristic of the closed string background represented by the bulk of the worldsheet CFT.
One may hope to regard this non-commutative algebra as a viable candidate which replaces the notion of classical spacetime geometry in string theory.

In particular, when the worldsheet CFT has a standard geometric interpretation, we hope that the non-commutative open string star algebra somehow contains the information of the massless closed string background fields such as the metric, Kalb-Ramond $B$-field, and dilaton.
In the past, there have been attempts to find on-shell closed string scattering states in open string field theory \cite{Freedman:1987fr,Gaiotto:2001ji}. 
Our goal here is related but different, since we aim to extract the classical closed string background fields from the star algebra.

The sentiment that open strings know about, or perhaps are more fundamental than, or are dual to closed strings is by now decades-old and there appear to be many different manifestations of this.
In addition to the original large-$N$ analysis by 't Hooft \cite{tHooft:1973alw} and the fact that it underlies the AdS/CFT correspondence \cite{Maldacena:1997re,Witten:1998qj,Gubser:1998bc}, inspiring works finding signatures of closed strings in various disguises in open string (field) theory include \cite{Gaiotto:2003yb,McGreevy:2003kb,Gaiotto:2003rm,Giddings:1986wp,Kawai:1985xq,Gopakumar:1998ki,Gopakumar:2022djw}.

With these broad motivations in mind, in this work, we consider the following situation.
Suppose we are given a smooth family of classical solutions $\Psi ( \lambda )$ of the open string field theory, where $\lambda = (\lambda^1 , \cdots , \lambda^N) \in X$ with $X$ being an $N$-dimensional subspace of the space of all classical solutions.
An example would be when the worldsheet conformal boundary condition around which we formulate the open string field theory admits exactly marginal deformations, and in that case the parameters $\lambda^i$ can be taken as the marginal boundary couplings (at least to the leading order) \cite{Fuchs:2007yy,Kiermaier:2007vu,Kiermaier:2007ba,Schnabl:2007az}.
We define a 2-form connection in the space $X$,\footnote{Our use of the word ``connection'' is not mathematically precise. By a connection we mean an anti-symmetric tensor field which has the usual gauge transformation law shown below in \eqref{eq:B_gauge_transform_intro}.\label{fn:connection}}
\begin{equation} \label{eq:B_definition}
    B_{ij} (\lambda) = \int \Psi * \frac{\partial \Psi}{\partial \lambda^i} * \frac{\partial \Psi}{\partial \lambda^j} - (i \leftrightarrow j) \,.
\end{equation}
The notations, which are standard in open string field theory, will be reviewed later.
At least in some simple cases, we would like to argue that there is a choice of $X$ which can be identified with the target spacetime of the given closed string background, and interpret $B_{ij}$ as the Kalb-Ramond $B$-field on $X$.
A valid choice of $X$ can be, for instance, a moduli space of a $D$-instanton.
Whether a given $X$ can be viewed as a moduli space of a $D$-instanton or not will in general change under various dualities. Relatedly, there can be multiple valid choices for $X$ which can be thought of as $D$-instanton moduli spaces in different duality frames.
We suggest the following:
\begin{center}
    \emph{When $X$ can be interpreted as a moduli space of a $D$-instanton,} \\
    \emph{$B_{ij} (\lambda)$ should coincide with the background Kalb-Ramond $B$-field.}
\end{center}
If true, it will provide us with new insights about how the open string star algebra contains the information of the massless closed string backgrounds when a geometric interpretation is available.

Independent of the physical meaning of \eqref{eq:B_definition}, it gives rise to new gauge-invariant quantities in open string field theory.
Open string field theory, as will be briefly reviewed, enjoys an infinite-dimensional gauge algebra.
It is a nontrivial task to come up with interesting observables invariant under these gauge transformations \cite{Hashimoto:2001sm,Gaiotto:2001ji,Ellwood:2008jh,Koyama:2020qfb}.
An infinitesimal gauge transformation of the string field $\Psi$ takes the form $\delta \Psi = Q\epsilon + \Psi * \epsilon - \epsilon * \Psi$ where $Q$ is the worldsheet BRST operator and $\epsilon$ is the gauge parameter.
Under such a gauge transformation, we show that \eqref{eq:B_definition} transforms as
\begin{equation} \label{eq:B_gauge_transform_intro}
    \delta B_{ij} = \frac{\partial \eta^{(\epsilon)}_j}{\partial \lambda^i} - \frac{\partial \eta^{(\epsilon)}_i}{\partial \lambda^j} \,,
\end{equation}
where
\begin{equation}
    \eta^{(\epsilon)}_i (\lambda) = \int \epsilon * \left( 
        \Psi * \Psi * \frac{\partial \Psi}{\partial \lambda^i} +\frac{\partial \Psi}{\partial \lambda^i} * \Psi * \Psi
    \right) \,.
\end{equation}
We may regard $\eta^{(\epsilon)}_i$ locally as a 1-form in $X$, and \eqref{eq:B_gauge_transform_intro} is the standard gauge transformation of a 2-form connection with the 1-form gauge parameter $\eta^{(\epsilon)}_i$.

It follows that the 3-form curvature associated with $B_{ij}$,
\begin{equation}
    H_{ijk}
    = \partial_i B_{jk} + \partial_j B_{ki} + \partial_k B_{ij} \,,
\end{equation}
and the holonomies around 2-cycles $\Sigma \subset X$,
\begin{equation}
    W(\Sigma) =
        \oint_\Sigma B
     \,, \quad B = \frac{1}{2} B_{ij} \mathrm{d}\lambda^i \mathrm{d}\lambda^j \,,
\end{equation}
are new observables in open string field theory which are invariant under the gauge transformation $\delta \Psi = Q\epsilon + \Psi * \epsilon - \epsilon * \Psi$.
An interesting aspect is that they are not associated with a single open string background, but rather with a family of classical solutions.
A better point of view is perhaps to regard them as physical observables associated with the \emph{closed string background} represented by the bulk of the worldsheet CFT.
To corroborate such an interpretation, we show that the 2-form connection $B_{ij}$, modulo gauge ambiguity, is independent of the choice of the reference conformal boundary condition on the worldsheet.
This also aligns with the proposal to identify $B_{ij}$ with the background Kalb-Ramond field in certain cases.

The expression \eqref{eq:B_definition} is somewhat reminiscent of the definition of Berry connection in quantum mechanics \cite{Berry:1984jv}.
There, one is given a family of quantum wavefunctions $\Psi (\vec{x};\lambda) = \langle \vec{x} |\Psi (\lambda)\rangle$ and the Berry connection is defined as 
\begin{equation} \label{eq:Berry}
A_i = \mathrm{i} \langle \Psi | \partial_i \Psi \rangle = \mathrm{i}\int \mathrm{d}^D \vec{x} ~~\overline{\Psi}\cdot\frac{\partial \Psi}{\partial \lambda^i}
\end{equation}
which is locally a 1-form in the parameter space.
Let us point out some obvious differences between \eqref{eq:B_definition} and \eqref{eq:Berry}.
In \eqref{eq:B_definition}, $\Psi$ is a classical open string field and not a quantum wavefunction.
It satisfies a reality condition and we do not have any complex numbers in \eqref{eq:B_definition}.
Also, the definition \eqref{eq:B_definition} is cubic in $\Psi$ and relatedly we have a 2-form connection instead of a 1-form connection.
The fact that only a 2-form connection is natural in open string field theory is related to the ghost number anomaly of the worldsheet CFT on a disk and this will be explained later.

In fact, our definition \eqref{eq:B_definition} is inspired by recent developments in condensed matter physics on what is known as ``higher Berry phase'' of quantum many-body systems \cite{Kitaev_Freed60,Kapustin:2020eby,PhysRevB.102.245113}.
In particular, we highlight that it has been observed by Ohyama and Ryu \cite{Ohyama:2023vus} that in a tensor network formulation of higher Berry phase, one encounters a mathematical structure which is analogous to the non-commutative algebra appearing in open string field theory.\footnote{For readers interested in detailed developments in the study of higher Berry phase and related topics in condensed matter physics and quantum field theory, we provide an extensive list of recent works: \cite{Kapustin:2020mkl,Cordova:2019jnf,Cordova:2019uob,Kapustin:2022apy,Choi:2022odr,Hsin:2022iug,Wen:2021gwc,Ohyama:2023suc,Beaudry:2023qyg,Qi:2023ysw,Spiegel:2023lhv,Shiozaki:2023xky,Sommer:2024dtb,Sommer:2024lzp,Artymowicz:2023erv,Geiko:2024cra,Manjunath:2024rxe,Shiozaki:2025pyo,Wen:2025xka,Ohyama:2024ytt,Prakash:2024yfr,Inamura:2024jke,Jones:2025khc,Manjunath:2026ezp,Drukker:2025dfm,Komargodski:2025jbu,Copetti:2025sym,Inamura:2026hjl,Ohyama:2026oay,Brennan:2026ira}.} 
We do not know if our $B_{ij}$ can be regarded as an actual Berry connection for an adiabatic process in open string field theory.

Over the past several decades, there has been steady development in open and/or closed string field theory, with or without supersymmetry \cite{Zwiebach:1992ie,Sen:2015uaa}. Some recent works are \cite{Stettinger:2024hkp,Stettinger:2024uus,Maccaferri:2025orz,Sen:2024npu,Mamade:2025htb,Mamade:2025jbs,Maccaferri:2025onc,Hull:2025mtb,Eniceicu:2022xvk,Sen:2025xaj,Mazel:2024alu,Mazel:2025fxj,Kudrna:2012re,Kudrna:2012um,Kudrna:2016ack,Kudrna:2014rya,Kudrna:2021rzd,Kudrna:2024ekn,Kojita:2016jwe,Kudrna:2018mxa,Erler:2019xof,Scheinpflug:2023osi,Schnabl:2023dbv,Scheinpflug:2023lfn,Schnabl:2024fdx,Scheinpflug:2025sqn,Firat:2021ukc,Firat:2023glo,Firat:2023suh,Erbin:2023hcs,Firat:2023gfn,Firat:2024ajp,Firat:2024kxq,Bernardes:2025zkj,Bernardes:2025zzu,Cho:2018nfn,Cho:2019anu,Cho:2023khj,Cho:2025coy,Cho:2023mhw,Kim:2024dnw,Frenkel:2025wko,Kim:2026kex,Berkovits:2024fpk,Ohmori:2017wtx,Okawa:2020llq,Okawa:2022mos}.
A more complete list of references can be found in \cite{Sen:2024nfd}.

The rest of the paper is organized as follows.
In Section \ref{sec:OSFT}, we review basics of bosonic open string field theory.
In Section \ref{sec:2-form}, we discuss properties of the 2-form connection \eqref{eq:B_definition} and explain why only 2-form connection naturally appears.
We then consider the 2-form connection for families of solutions to open string field theory coming from the marginal deformations of the worldsheet boundary condition.
We also briefly review the previous related work \cite{Choi:2025ebk}.
In Section \ref{sec:sigma_models}, we point out that the moduli space of conformal boundary conditions of a two-dimensional CFT is always a natural target space for a sigma model, as it is equipped with a metric and a $B$-field. 
Both the metric and $B$-field are defined in terms of boundary correlation functions.
We study examples where such a sigma model is equivalent to the original CFT.
Finally, we conclude in Section \ref{sec:discussion} with discussion and outlook.

\section{Open string field theory} \label{sec:OSFT}

We briefly recall the formulation of bosonic open string field theory \cite{Witten:1985cc}.
There are no new materials in this section.
A conceptually simple way to proceed is to first start from an abstract non-commutative algebra $\mathcal{A}$.
In practice, the non-commutative algebra should arise from a worldsheet boundary CFT (with various subtleties) but a priori there seems to be no reason to restrict only to such cases.

The non-commutative algebra $\mathcal{A}$ must possess the following additional structures.
First, it comes with an integer grading, $\mathcal{A} = \bigoplus_{n\in\mathbb{Z}} \mathcal{A}_n$. 
The integer $n$ is called the ``ghost number,'' and this will be the usual ghost number if we start from a worldsheet CFT.
The ghost number of an element $\Psi \in \mathcal{A}$ will be denoted by $|\Psi|$.
The multiplication is called the ``star product.'' 
Given two elements $\Psi_1$ and $\Psi_2$ of the algebra, their star product is denoted by $\Psi_1 * \Psi_2$.
In general, the star product need not be commutative, $\Psi_1 * \Psi_2 \neq \Psi_2 * \Psi_1$, but it must be associative. 
We have $|\Psi_1 * \Psi_2| = |\Psi_1 | + |\Psi_2 |$.

Next, the algebra $\mathcal{A}$ comes with a derivation $Q$ of degree 1.
It is nilpotent, $Q^2 = 0$, and it satisfies the Leibniz rule $Q(\Psi_1 * \Psi_2 ) = (Q\Psi_1 ) * \Psi_2  + (-1)^{|\Psi_1|} \Psi_1 * (Q\Psi_2  )$.
When we have an underlying worldsheet CFT, $Q$ is the worldsheet BRST operator \cite{Siegel:1984ap,Siegel:1984ogw,Siegel:1984xd}.
With the derivation $Q$, $\mathcal{A}$ is now a differential graded algebra.

We need one more additional structure, which is the trace on the algebra $\mathcal{A}$.
The trace is called the ``integration'' and is denoted by $\int$.
It is a linear functional, $\int : \mathcal{A} \rightarrow \mathbb{C}$, and it must satisfy the following conditions.
First, we can ``integrate by parts,'' $\int Q \Psi = 0$ for all $\Psi \in \mathcal{A}$.
Next, $\int$ is indeed a trace, $\int \Psi_1 * \Psi_2 = (-1)^{|\Psi_1|\cdot|\Psi_2|} \int \Psi_2 * \Psi_1$.
Finally, the integration is concentrated at ghost number 3, $\int \Psi = 0$ if $|\Psi| \neq 3$.

With these ingredients, we are ready to formulate the open string field theory.
The theory is defined by the action
\begin{equation} \label{eq:osft_action}
    I = \int \left( 
        \Psi * Q\Psi + \frac{2}{3} \Psi * \Psi * \Psi
    \right) \,.
\end{equation}
Here, the ``open string field'' $\Psi$ is an element of the algebra $\mathcal{A}$ and it is the degree of freedom of the theory.
We are only interested in the classical aspects of the theory in this paper.
Classically, $\Psi$ is restricted to have ghost number 1.
Note that the integrand in \eqref{eq:osft_action} then has ghost number 3 and this is consistent with the requirement that the integration is concentrated at ghost number 3.

The action \eqref{eq:osft_action} is invariant under the gauge transformation
\begin{equation}
    \delta \Psi = Q \epsilon + \Psi * \epsilon - \epsilon * \Psi \,,
\end{equation}
where the gauge parameter $\epsilon$ is an arbitrary element of $\mathcal{A}$ with ghost number 0.
The gauge transformations form an algebra
\begin{equation}
    [\delta_{\epsilon_1} , \delta_{\epsilon_2}] \Psi = \delta_{[\epsilon_1 , \epsilon_2]} \Psi \,,
\end{equation}
where $[\epsilon_1 , \epsilon_2] \equiv \epsilon_1 * \epsilon_2 - \epsilon_2 * \epsilon_1$.
The form of the action and the gauge transformation is formally identical to that of the Chern-Simons theory based on a non-abelian gauge group \cite{Witten:1988hf}.
In fact, an example of an algebra obeying above axioms is the space of matrix-valued differential forms on a 3-dimensional manifold \cite{Witten:1985cc}, with the product given by the wedge product and the derivation given by the exterior derivative.
In that case, the action \eqref{eq:osft_action} reduces to that of a Chern-Simons theory (at least locally).
See also \cite{Witten:1992fb}.

Finally, the classical equation of motion which is derived from the action \eqref{eq:osft_action} in the usual way is
\begin{equation} \label{eq:EOM}
    Q\Psi + \Psi * \Psi = 0 \,.
\end{equation}
Writing $F \equiv Q\Psi + \Psi * \Psi$, we have
\begin{equation}
    \delta F = F * \epsilon - \epsilon * F \,.
\end{equation}
That is, the equation of motion transforms covariantly under the gauge transformation.

\begin{figure}[t!]
\centering
\includegraphics[width=.8\textwidth]{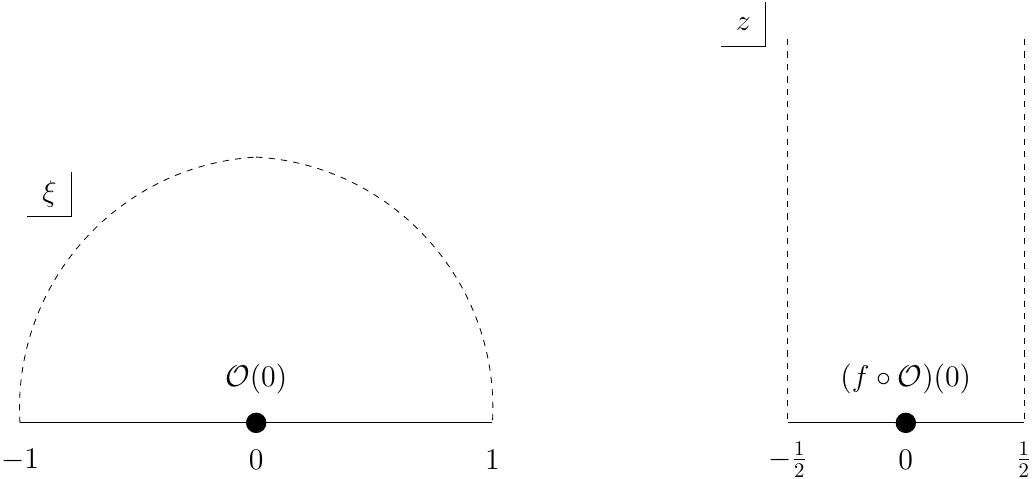}
\caption{A typical element of $\mathcal{A}$ is prepared by a path integral on a unit half-disk with operator insertions. It is convenient to map to the sliver frame shown in right.
In both cases, along the real line we impose the reference boundary condition $\mathcal{B}$. Along the dotted lines the boundary condition is unspecified.}\label{fig:sliver}
\end{figure}

Let us explain how the above structure arises from the worldsheet CFT.\footnote{It seems difficult to extract a non-commutative algebra from the worldsheet CFT in a mathematically rigorous way. One important issue is the so-called midpoint anomaly \cite{Bars:2002bt,Erler:2003eq} which may affect the associativity of the algebra.}
A helpful reference is \cite{Erler:2019vhl}.
Suppose we are given a consistent (bosonic) worldsheet CFT which is composed of the matter CFT of central charge $26$ and $bc$ ghosts.
We choose a conformal boundary condition $\mathcal{B}$ of the worldsheet CFT. 
We will construct a non-commutative algebra $\mathcal{A}$ whose underlying vector space is the Hilbert space $\mathcal{H}_{\mathcal{B}}$ of the worldsheet CFT on an interval with the boundary condition $\mathcal{B}$ imposed on both ends.\footnote{To be more pedantic, the non-commutative algebra $\mathcal{A}$ will contain the Hilbert space $\mathcal{H}_\mathcal{B}$ as a subspace and in general will be larger than $\mathcal{H}_\mathcal{B}$. For instance, allowed string field configurations may include infinite linear combinations of states in $\mathcal{H}_{\mathcal{B}}$ which may not be normalizable. One possibility could be to consider the ``rigged Hilbert space'' which contains $\mathcal{H}_{\mathcal{B}}$ as a subspace. (A recent appearance of the rigged Hilbert space in the context of boundary CFT can be found in \cite{Popov:2025cha}.)}
We do not impose Virasoro constraints and the states need not be BRST closed.

It is helpful to visualize how typical elements of $\mathcal{A}$ are obtained physically through a path integral.
For instance, take a primary state $\ket{\mathcal{O}}$.
This state is prepared through a Euclidean path integral on a unit half-disk with the corresponding boundary primary operator $\mathcal{O}(0)$ inserted at the origin, as shown in the left half of Figure \ref{fig:sliver}.
For practical computations, it is common to map the unit half disk to a half-infinite strip through a conformal transformation \cite{Schnabl:2005gv}
\begin{equation} \label{eq:sliver}
    z = f(\xi) = \frac{2}{\pi} \mathrm{tan}^{-1} \xi \,.
\end{equation}
This is called the ``sliver frame.''
See the right half of Figure \ref{fig:sliver}.
The primary operator $\mathcal{O}$ transforms accordingly.
More general elements of $\mathcal{A}$ can be prepared by a similar path integral where one decorates the half disk, or the half-infinite strip in the sliver frame, with various other operator insertions.
This can include both bulk and boundary operators, and we may also replace segments of the reference conformal boundary condition $B$ with some other boundary condition $B'$, with boundary-condition-changing operators inserted at the interfaces between $B$ and $B'$.
See, for instance, \cite{Erler:2019fye}.

\begin{figure}[t!]
\centering
\includegraphics[width=.9\textwidth]{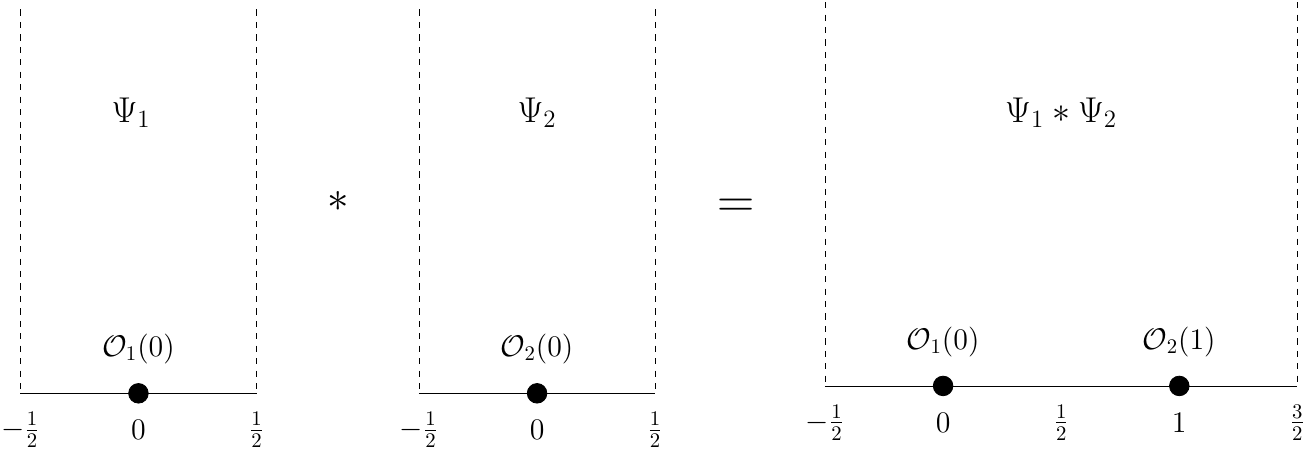}
\caption{In the sliver frame, the star product $\Psi_1 * \Psi_2$ is obtained by a path integral on a half-infinite strip which is obtained by gluing the two half-infinite strips for $\Psi_1$ and $\Psi_2$. More complicated operator insertions are also possible.}\label{fig:product}
\end{figure}

The star product and integration of the string fields are defined as follows.
One of the intriguing insights from \cite{Witten:1985cc} was to define a non-commutative product of two open strings by ``gluing'' the left-half of one of the strings to the right-half of the other string, producing a new open string.
In the sliver frame, the left-half of the string corresponds to the vertical line $z = -1/2$ and the right-half to the vertical line $z= + 1/2$.
Given two open string fields $\Psi_1$ and $\Psi_2$ represented by path integrals on half-infinite strips, the star product $\Psi_1 * \Psi_2$ is represented by a path integral on a new, slightly thicker half-infinite strip obtained by gluing the two previous strips.
This is shown in Figure \ref{fig:product}.
Finally, the integration $\int \Psi$ is defined by the correlation function of the worldsheet CFT on a surface obtained by gluing left and right halves of $\Psi$.
In the sliver frame, this produces a correlation function on a half-infinite cylinder, which is conformally equivalent to a disk or the upper half-plane.

\section{2-form connection} \label{sec:2-form}

\subsection{Definition and basic properties}

Suppose we have a smooth family of parameterized solutions $\Psi (\lambda)$ to the equation of motion \eqref{eq:EOM}.
That is,
\begin{equation} \label{eq:EOM_lambda}
    Q\Psi (\lambda ) + \Psi (\lambda) * \Psi (\lambda) = 0 ~~~\text{for all $\lambda = (\lambda^1 , \cdots , \lambda^N) \in X$} \,.
\end{equation}
Here, $X$ is an $N$-dimensional subspace of the space of all classical solutions.
An example will be discussed shortly.
Let us reiterate the definition of our 2-form connection in \eqref{eq:B_definition},\footnote{As mentioned in Footnote \ref{fn:connection}, our use of the word ``connection'' is slightly imprecise. At the moment, we only know how to define $B_{ij}$ on a local patch on $X$. We anticipate that our $B_{ij}$ can be promoted to an actual connection on a gerbe, perhaps using the ideas similar to those in \cite{Ohyama:2023vus,Ohyama:2024jsg}.}
\begin{equation} \label{eq:B_definition_2}
    B_{ij} (\lambda) = \int \Psi * \frac{\partial \Psi}{\partial \lambda^i} * \frac{\partial \Psi}{\partial \lambda^j} - (i \leftrightarrow j) \,.
\end{equation}
This is defined abstractly using the non-commutative star algebra $\mathcal{A}$ independent of whether there is an underlying worldsheet CFT or not.
Note that the non-commutative nature of the star product is crucial for \eqref{eq:B_definition_2} to be non-vanishing.

An important consistency check is to analyze the gauge transformation of \eqref{eq:B_definition_2}.
We allow the gauge parameter $\epsilon$ to be also $\lambda$-dependent and let us not impose the equation of motion \eqref{eq:EOM_lambda} until the very end.
Under the gauge transformation of the open string field,
\begin{equation} \label{eq:psi_transform}
    \delta \Psi (\lambda) = Q \epsilon (\lambda) + \Psi (\lambda) * \epsilon (\lambda) - \epsilon (\lambda) * \Psi (\lambda) \,,
\end{equation}
the 2-form connection \eqref{eq:B_definition_2} transforms as
\begin{align}
\begin{split}
    \delta B_{ij} &= \partial_i  \left[ \int \left(
        (Q\epsilon) * \Psi *  \partial_j \Psi + \epsilon * \partial_j \Psi^3 \right)
    \right] \\
    &~~~~+ \int \partial_i \epsilon \Bigg[
        \partial_j \Psi * \left( 
            Q \Psi + \Psi^2
        \right) - \left(\partial_j (Q\Psi + \Psi^2) \right) * \Psi \\
        &~~~~~~~~~~~~~~~~~~~~+ 2 (Q\Psi + \Psi^2) * \partial_j \Psi - 2\Psi * \left(\partial_j (Q\Psi + \Psi^2) \right)
    \Bigg] - (i \leftrightarrow j) \,.
\end{split}
\end{align}
This is derived using the axioms for the non-commutative algebra $\mathcal{A}$ which was explained in the previous section.
Imposing the equation of motion $Q\Psi + \Psi^2 = 0$ and then simplifying, we obtain
\begin{equation}
    \delta B_{ij} = \partial_i \left[ \int
        \epsilon * \left( \Psi^2 * \partial_j \Psi + (\partial_j \Psi) * \Psi^2 \right)
    \right] - (i \leftrightarrow j) \,,
\end{equation}
which takes the form of a usual gauge transformation of a 2-form connection.
That is, if we write $B = \frac{1}{2} B_{ij} \mathrm{d}\lambda^i \mathrm{d}\lambda^j$, we have
\begin{equation} \label{eq:B_gauge_transform}
    \delta B = \mathrm{d}\eta^{(\epsilon)} \,,
\end{equation}
where
\begin{equation}
    \eta^{(\epsilon)} = \eta^{(\epsilon)}_i (\lambda) \mathrm{d}\lambda^i \,, \quad \eta^{(\epsilon)}_i (\lambda) =  \int
        \epsilon * \left( \Psi^2 * \partial_i \Psi + (\partial_i \Psi) * \Psi^2 \right) \,.
\end{equation}
Therefore, under \eqref{eq:psi_transform}, $B_{ij}$ indeed transforms as a 2-form connection, with the 1-form gauge parameter given by $\eta^{(\epsilon)}_i$.
We comment that if we restrict the gauge parameter $\epsilon$ to be independent of $\lambda$, then we can generalize \eqref{eq:B_definition_2} to include off-shell string field configurations (that is, we do not impose $Q\Psi + \Psi^2=0$), and still obtain a valid 2-form connection.

The immediate consequence of \eqref{eq:B_gauge_transform} is that the 3-form curvature
\begin{equation} \label{eq:curvature}
    H = \mathrm{d}B \,, \quad H = \frac{1}{3!} H_{ijk} \mathrm{d}\lambda^i \mathrm{d}\lambda^j \mathrm{d}\lambda^k \,,
\end{equation}
with 
$H_{ijk} = \partial_i B_{jk} + \partial_j B_{ki} + \partial_k B_{ij}$, is invariant under the gauge transformation of the open string field \eqref{eq:psi_transform}.
Explicitly, we have
\begin{equation} \label{eq:curvature_explicit}
    H_{ijk} = 3 \int \partial_i \Psi * \left( 
        \partial_j \Psi * \partial_k \Psi - \partial_k \Psi * \partial_j \Psi
    \right) \,.
\end{equation}
Furthermore, for arbitrary 2-cycles $\Sigma \subset X$, the holonomy
\begin{equation} \label{eq:holonomy}
    W(\Sigma) = \oint_{\Sigma} B
\end{equation}
is also invariant under \eqref{eq:psi_transform}.
Therefore, the 2-form connection \eqref{eq:B_definition_2} gives rise to new kinds of gauge-invariant observables in open string field theory, namely the 3-form curvature \eqref{eq:curvature} and the holonomies \eqref{eq:holonomy} around different 2-cycles in $X$.

Let us make an important observation.
We will show that the 2-form connection \eqref{eq:B_definition_2} is independent of the choice of the reference conformal boundary condition on the worldsheet, modulo gauge ambiguity.
Suppose we pick a different reference conformal boundary condition on the worldsheet and formulate the open string field theory around this new conformal boundary condition.
Let us denote the new non-commutative algebra by $\mathcal{A}'$, integration by $\int'$, and open string field by $\Psi'$.
It was claimed in \cite{Erler:2019fye}, generalizing earlier works \cite{Sen:1990hh,Sen:1992pw,Sen:1993mh,Sen:1993kb}, that the open string field theory formulated with the new choice of the reference conformal boundary condition is equivalent to the original one by a field redefinition
\begin{equation} \label{eq:field_redefinition}
    \Psi = \Psi_0 + f (\Psi') \,.
\end{equation}
Here, $\Psi_0$ is a classical solution to the equation of motion in the original open string field theory which describes the $D$-brane configuration corresponding to the new reference conformal boundary condition, and $f: \mathcal{A}' \rightarrow \mathcal{A}$ is a linear map which is invertible modulo gauge ambiguity.
The linear map $f$ has been argued to possess following properties:
\begin{equation} \label{eq:f_properties}
    f(\Psi_1' * \Psi_2' ) = f(\Psi_1' ) * f(\Psi_2' ) \,,\quad
    \int f(\Psi') = \int' \Psi' \,,\quad
    f(Q\Psi') = Q_{\Psi_0} f(\Psi') \,,
\end{equation}
where $Q_{\Psi_0} \Psi \equiv Q\Psi + \Psi_0 * \Psi - (-1)^{|\Psi|} \Psi * \Psi_0$ for any $\Psi \in \mathcal{A}$.

Now, take the family of classical solutions $\Psi' (\lambda)$, $\lambda \in X$, of the open string field theory formulated around the new reference conformal boundary condition.
They are related to the original family of solutions $\Psi (\lambda)$ by the field redefinition \eqref{eq:field_redefinition}, $\Psi (\lambda) = \Psi_0 + f(\Psi' (\lambda))$ for all $\lambda \in X$.
$\Psi_0$ is independent of $\lambda$.
Denote the 2-form connection corresponding to the family $\Psi' (\lambda)$ by $B'_{ij} (\lambda)$. 
That is,
\begin{equation}
    B'_{ij} (\lambda) = \int' \Psi' * \frac{\partial \Psi'}{\partial \lambda^i} * \frac{\partial \Psi'}{\partial \lambda^j} - (i \leftrightarrow j)  \,.
\end{equation}
Let us compare this with the original 2-form connection $B_{ij} (\lambda)$.
Using \eqref{eq:field_redefinition}, \eqref{eq:f_properties} and the fact that $f$ is a linear map, we have
\begin{align}
\begin{split}
    B_{ij} (\lambda) &= \int \Psi * \frac{\partial \Psi}{\partial \lambda^i} * \frac{\partial \Psi}{\partial \lambda^j} - (i \leftrightarrow j) \\
    &= \int \left( 
        \Psi_0 + f ( \Psi' )
    \right) * \frac{\partial f(\Psi')}{\partial \lambda^i} * \frac{\partial f(\Psi')}{\partial \lambda^j} - (i \leftrightarrow j) \\
    &= \int \left[ f\left( 
        \Psi' * \frac{\partial \Psi'}{\partial \lambda^i} * \frac{\partial \Psi'}{\partial \lambda^j}
    \right) + \Psi_0 * \frac{\partial f(\Psi')}{\partial \lambda^i} * \frac{\partial f(\Psi')}{\partial \lambda^j} \right] - (i \leftrightarrow j) \\
    &= B'_{ij}(\lambda) + \partial_i \eta_j^{(\Psi_0 )} - \partial_j \eta_i^{(\Psi_0 )}\,.
\end{split}
\end{align}
We learn that $B_{ij} (\lambda)$ and $B'_{ij} (\lambda)$ are equivalent up to a gauge transformation by a 1-form gauge parameter
\begin{equation}
    \eta^{(\Psi_0 )} = \eta^{(\Psi_0 )}_i (\lambda) \mathrm{d}\lambda^i \,, \quad \eta^{(\Psi_0 )}_i (\lambda) = \int \Psi_0 * f(\Psi') * \partial_i f(\Psi' ) \,.
\end{equation}
We have shown that the 2-form connection $B_{ij}$, modulo gauge ambiguity, does not depend on the choice of the reference conformal boundary condition on the worldsheet.
The 3-form curvature \eqref{eq:curvature} and holonomies \eqref{eq:holonomy} are therefore to be regarded as gauge-invariant observables associated with the fixed closed string background represented by the bulk of the worldsheet CFT.

A natural question is whether one can define an arbitrary $n$-form connection on $X$.
For instance, one may wonder whether
\begin{equation}
    B_{i_1 i_2 \cdots i_n} = \int \Psi * \frac{\partial \Psi}{\partial \lambda^{i_1}} * \cdots * \frac{\partial \Psi}{\partial \lambda^{i_n}} + \cdots
\end{equation}
can be defined for more general $n\neq 2$.
It is easy to realize that $B_{i_1 i_2 \cdots i_n}$ vanishes unless $n=2$.
This is because the classical string field $\Psi$ has ghost number 1 and the integration $\int$ vanishes unless the total ghost number of the integrand is 3.
On the worldsheet, this condition arises from the ghost number anomaly of disk correlation functions.
We see that the definition \eqref{eq:B_definition_2} does not generalize to an arbitrary $n$-form with $n\neq 2$ in a straightforward way.
This can potentially be circumvented by inserting additional fixed string fields of arbitrary ghost number and/or the BRST operator $Q$.

Finally, we mention that the overall normalization of the 2-form connection \eqref{eq:B_definition_2} is meaningful.
The equation of motion of the open string field theory, $Q\Psi + \Psi^2 =0$, being non-linear, does not allow the on-shell string fields to be arbitrarily rescaled.
Therefore, one can meaningfully ask, for instance, whether the 3-form curvature \eqref{eq:curvature} has quantized periods or not.\footnote{In condensed matter physics, there are rigorous approaches to studying higher Berry phase where it has been shown that the higher Berry curvature has quantized periods \cite{Artymowicz:2023erv}.}

\subsection{Marginal solutions} \label{sec:marginal}

As was explained, to formulate the open string field theory starting from a worldsheet CFT, one chooses a reference conformal boundary condition of the worldsheet CFT.
This reference boundary condition corresponds to the zero string field, $\Psi = 0$, which trivially solves the equation of motion.
All the other conformal boundary conditions of the worldsheet CFT appear as nontrivial solutions to the open string field theory equation of motion \cite{Erler:2014eqa,Kiermaier:2010cf,Erler:2019fye,Erler:2019vhl}.
An extensively studied class of examples is when the reference conformal boundary condition admits exactly marginal boundary operators \cite{Fuchs:2007yy,Kiermaier:2007vu,Kiermaier:2007ba,Schnabl:2007az}.
In this case, deformation of the reference boundary condition by the marginal operators gives us a family of classical solutions to the open string field equation of motion, parameterized by the boundary marginal couplings of the worldsheet CFT.
Here, we discuss the 2-form connection $B_{ij}$ for such a family.

Let $V_{i}$, $i = 1, \cdots, N$, be exactly marginal boundary operators of the worldsheet matter CFT living on the reference conformal boundary condition around which we have formulated the open string field theory.
(See \cite{Recknagel:1998ih,Callan:1994ub,Gaberdiel:2008fn,Gaberdiel:2001zq} for various examples.)
We perturb the reference boundary condition of the worldsheet CFT by a linear combination $\sum_{i=1}^N \lambda^i V_i$ where $\lambda^i$'s are the marginal coupling constants.
Following \cite{Kiermaier:2007ba,Schnabl:2007az}, we seek for a perturbative expansion for the solutions to the equation of motion $Q\Psi + \Psi^2 = 0$ describing the deformed worldsheet boundary conditions, of the form
\begin{equation} \label{eq:psi_pert}
    \Psi (\lambda) =\sum_{i=1}^N \lambda^i \Psi^{(1)}_i + \sum_{i,j=1}^N \lambda^i \lambda^j \Psi^{(2)}_{ij} + \cdots \,.
\end{equation}
The equation of motion, order by order in marginal couplings, leads to
\begin{align}
\begin{split}
    Q\Psi_i^{(1)} &= 0 \,,\\
    Q\Psi_{ij}^{(2)} &= - \frac{1}{2} \left(\Psi_i^{(1)} * \Psi_j^{(1)} + \Psi_j^{(1)} * \Psi_i^{(1)} \right) \,,
\end{split}
\end{align}
and so on.
We will content ourselves with keeping track of the expansion \eqref{eq:psi_pert} only up to the terms that are second order in $\lambda^i$.

The first order equation of motion requires $\Psi_i^{(1)}$ to be BRST closed.
It is clear that we have to take
\begin{equation}
    \Psi_i^{(1)} = cV_i \,,
\end{equation}
with $c$ being the $c$ ghost.
For the second order terms $\Psi_{ij}^{(2)}$, we adopt the result of \cite{Kiermaier:2007ba,Schnabl:2007az}.
It is convenient to characterize it by its product with a test operator $\phi$,
\begin{equation} \label{eq:KORZ}
    \langle \phi , \Psi_{ij}^{(2)} \rangle \equiv \int \phi * \Psi_{ij}^{(2)} = \frac{1}{2}\int_{0}^1 \mathrm{d}t ~ \langle (f \circ \phi) (0) cV_{i}(1) \mathcal{B} cV_{j}(1+t) \rangle_{\mathcal{W}_{1+t}} + (i \leftrightarrow j) \,.
\end{equation}
We are following the notations of \cite{Kiermaier:2007ba}.
On the right-hand side, $f$ is the conformal transformation to the sliver frame given in \eqref{eq:sliver}, and $\mathcal{B} = \int \frac{dz}{2\pi i} b(z)$ is the integral of the $b$ ghost along the contour that runs anti-parallel to the imaginary axis (using doubling trick) with the real part of $z$ chosen to be anywhere between $1 < \mathrm{Re}(z) < 1+t$.
The details will not be important for us.
$\mathcal{W}_{\alpha}$ is a surface obtained from a half-infinite strip, $-\frac{1}{2} \leq\mathrm{Re}(z) \leq \frac{1}{2} + \alpha$, $\mathrm{Im}(z) \geq 0$, by gluing the two sides, $\mathrm{Re}(z) = -\frac{1}{2}$ and $\mathrm{Re}(z) = \frac{1}{2} + \alpha$.
Here, $\alpha$ is any non-negative real number.
We have $\alpha = 1+t$ above.

There is an important subtlety we will gloss over.
In \eqref{eq:KORZ}, the integrand on the right-hand side will in general diverge at $t=0$ due to the singularity coming from the collision of the two marginal operators $V_i$ and $V_j$.
Such a divergence can be carefully regularized following \cite{Kiermaier:2007ba}.
We will be agnostic about this and aim to only have a rough intuition of what the 2-form connection $B_{ij}$ computes on the worldsheet using the perturbative solution.
Or rather, let us focus on the curvature $H_{ijk}$ in \eqref{eq:curvature_explicit} which is gauge-invariant.

We expand\footnote{In \eqref{eq:H_expansion}, the coefficients of higher-order terms such as $H_{ijk;a}^{(1)}$ are to be thought of as covariant derivatives of $H_{ijk}$ at $\lambda^a = 0$. However, at the moment we do not know how to properly define a metric and the corresponding affine connection on $X$. Practically, the expansion \eqref{eq:H_expansion} comes from inserting the perturbative expansion \eqref{eq:psi_pert} of the open string field into the definition \eqref{eq:curvature_explicit} of the curvature $H_{ijk}$.}
\begin{equation}\label{eq:H_expansion}
    H_{ijk} = H^{(0)}_{ijk} + \sum_{a=1}^N \lambda^a H^{(1)}_{ijk;a} + \cdots \,.
\end{equation}
From the perturbative expansion \eqref{eq:psi_pert}, we obtain
\begin{align}
\begin{split}
    H_{ijk}^{(0)} &= 3 \int \left( 
        \Psi_{i}^{(1)} * \Psi_{j}^{(1)} * \Psi_{k}^{(1)} - \Psi_{i}^{(1)} * \Psi_{k}^{(1)} * \Psi_{j}^{(1)}
    \right) \,,\\
    H_{ijk;a}^{(1)} &= 6 \int \Bigg( 
        \Psi_{ia}^{(2)} * \Psi_j^{(1)} * \Psi_k^{(1)} - \Psi_{ia}^{(2)} * \Psi_k^{(1)} * \Psi_j^{(1)} \\
        &~~~~~~ +\Psi_{i}^{(1)} * \Psi_{ja}^{(2)} * \Psi_{k}^{(1)} - \Psi_{i}^{(1)} * \Psi_{ka}^{(2)} * \Psi_{j}^{(1)} \\
        &~~~~~~ +\Psi_{i}^{(1)} * \Psi_{j}^{(1)} * \Psi_{ka}^{(2)} - \Psi_{i}^{(1)} * \Psi_{k}^{(1)} * \Psi_{ja}^{(2)}
    \Bigg) \,.
\end{split}
\end{align}
Let us first take a look at the zeroth order term.
We have
\begin{equation}
    \int \Psi_{i}^{(1)} * \Psi_{j}^{(1)} * \Psi_{k}^{(1)} = \langle cV_i (0) cV_j (1) cV_k (2) \rangle_{\mathcal{W}_2} = f_{ijk}
\end{equation}
where $f_{ijk}$ is the OPE coefficient of the three marginal operators $V_i$, $V_j$, $V_k$ of the matter CFT.\footnote{Our normalization convention for the OPE coefficient is such that, on a unit disk $\zeta = re^{i\theta}$,
\begin{equation}
    \langle V_{i} (\theta_1 ) V_{j} (\theta_2 ) V_k (\theta_3) \rangle = f_{ijk} \left[
        2\mathrm{sin} \left( \frac{ \theta_2 - \theta_1}{2} \right)
    \right]^{-1} \left[
        2\mathrm{sin} \left( \frac{ \theta_3 - \theta_2}{2} \right)
    \right]^{-1} \left[
        2\mathrm{sin} \left( \frac{ \theta_1 - \theta_3}{2} \right)
    \right]^{-1} \,,
\end{equation}
where $0 \leq \theta_1 < \theta_2 < \theta_3 <2\pi$.
}
Exact marginality requires that $f_{ijk}$ is anti-symmetric in its indices.\footnote{Sometimes it is said that for the operators $V_i$ to be exactly marginal, a necessary condition is to have $f_{ijk} =0$ for all $i,j,k$. This is not correct. For instance, a common example of boundary exactly marginal operators is the set of holomorphic current operators, restricted to a boundary, in a Wess-Zumino-Witten model \cite{Recknagel:1998ih}. There, the OPE coefficient of current operators is proportional to the structure constant of the underlying Lie algebra.}
Therefore, we obtain
\begin{equation}
    H_{ijk}^{(0)} = 6f_{ijk} \,.
\end{equation}

Next, for the first order term $H_{ijk;a}^{(1)}$, we need to compute terms of the form
\begin{equation} \label{eq:H1}
    \int \Psi_i^{(1)} * \Psi_j^{(1)} * \Psi_{ka}^{(2)} = \frac{1}{2}\int_0^1 \mathrm{d} t ~ \langle cV_i (0) cV_j (1) cV_k (2) \mathcal{B} cV_a (2+t) \rangle_{\mathcal{W}_{2+t}} + (k \leftrightarrow a) \,.
\end{equation}
Similar to before, $\mathcal{B} = \int \frac{dz}{2\pi i} b(z)$ is integrated along a contour anti-parallel to the imaginary axis with the real part of $z$ anywhere in the range $2<\mathrm{Re}(z)<2+t$, and doubling trick is used.
We can factorize the correlation function in the integrand into the ghost part and the matter CFT part,
\begin{align}
\begin{split}
    &~~~~~~\langle cV_i (0) cV_j (1) cV_k (2) \mathcal{B} cV_a (2+t) \rangle_{\mathcal{W}_{2+t}} \\
    &= \langle c (0) c (1) c (2) \mathcal{B} c (2+t) \rangle_{\mathcal{W}_{2+t}}^{\text{ghost}} \times \langle V_i (0) V_j (1) V_k (2) V_a (2+t) \rangle_{\mathcal{W}_{2+t}}^{\text{matter}} \\
    &= X(t) \langle V_i (0) V_j (1) V_k (2) V_a (2+t) \rangle_{\mathcal{W}_{2+t}}^{\text{matter}} \,.
\end{split}
\end{align}
We have written $X(t) \equiv \langle c (0) c (1) c (2) \mathcal{B} c (2+t) \rangle_{\mathcal{W}_{2+t}}^{\text{ghost}}$ which can be explicitly evaluated \cite{Schnabl:2005gv,Okawa:2006sn}.
For the matter part, we may perform a conformal transformation to the upper half-plane,
\begin{equation}
    \xi (z ) = \frac{1}{2} \mathrm{tan} \left( 
        \frac{\pi(z-1)}{3+t}
    \right) \,,
\end{equation}
and obtain
\begin{align}
\begin{split}
    &~~~~~~\langle V_i (0) V_j (1) V_k (2) V_a (2+t) \rangle_{\mathcal{W}_{2+t}}^{\text{matter}} \\
    &= \left( 
        \frac{\pi}{2}\frac{1}{3+t}
    \right)^4 \mathrm{sec}^4 \left(
        \frac{\pi}{3+t}
    \right) \mathrm{sec}^2 \left( 
        \frac{\pi (1+t)}{3+t}
    \right) \\
    &~~~~~~~~~~~\times\Bigg\langle 
        V_i \left( 
            -\frac{1}{2}\mathrm{tan}\left(
                \frac{\pi}{3+t}
            \right)
        \right)V_j \left( 0
        \right)V_k \left( 
            \frac{1}{2}\mathrm{tan}\left(
                \frac{\pi}{3+t}
            \right)
        \right)V_a \left( 
            \frac{1}{2}\mathrm{tan}\left(
                \frac{\pi (1+t)}{3+t}
            \right)
        \right)
    \Bigg\rangle^{\text{matter}} \,.
\end{split}
\end{align}
We use $\langle \cdots \rangle$ without any subscript to denote correlation functions on the upper half-plane.

Plugging these ingredients into \eqref{eq:H1}, we have
\begin{align}
\begin{split}
    &~~~~~\int \Psi_i^{(1)} * \Psi_j^{(1)} * \Psi_{ka}^{(2)} \\
    &= \frac{1}{2}\int_0^1 \mathrm{d}t X' (t) \Bigg\langle 
        V_i \left( 
            -\frac{1}{2}\mathrm{tan}\left(
                \frac{\pi}{3+t}
            \right)
        \right)V_j \left( 0
        \right)V_k \left( 
            \frac{1}{2}\mathrm{tan}\left(
                \frac{\pi}{3+t}
            \right)
        \right)V_a \left( 
            \frac{1}{2}\mathrm{tan}\left(
                \frac{\pi (1+t)}{3+t}
            \right)
        \right)
    \Bigg\rangle^{\text{matter}} \\
    &~~~~~~~~~+(k\leftrightarrow a)\,,
\end{split}
\end{align}
where
\begin{equation}
    X' (t) \equiv \left( 
        \frac{\pi}{2}\frac{1}{3+t}
    \right)^4 \mathrm{sec}^4 \left(
        \frac{\pi}{3+t}
    \right) \mathrm{sec}^2 \left( 
        \frac{\pi (1+t)}{3+t}
    \right) \times \langle c (0) c (1) c (2) \mathcal{B} c (2+t) \rangle_{\mathcal{W}_{2+t}}^{\text{ghost}} \,.
\end{equation}
This then allows us to evaluate $H^{(1)}_{ijk;a}$ if the 4-point correlation functions of the operators $V_i$ are known.

Order by order in perturbative expansion, one encounters similar structures.
That is, the 3-form curvature $H_{ijk} (\lambda)$ at a given order in the perturbative expansion with respect to $\lambda$ is determined by integrated correlation functions of the marginal operators $V_i$ of the worldsheet matter CFT.
An important question is whether such an expansion can be summed up into an intuitive expression.
Perhaps a more promising way to proceed is to utilize the exact expression for the family of on-shell open string fields $\Psi (\lambda)$ given in terms of boundary-condition-changing operators \cite{Erler:2019fye}.
We will not explicitly pursue this here, but below we give a related discussion.

\subsection{Another version} \label{sec:another_version}

In a recent work \cite{Choi:2025ebk}, it was suggested that the space of conformal boundary conditions of a two-dimensional CFT is always equipped with a 2-form connection determined in terms of correlation functions of boundary-condition-changing (bcc) operators.
Let us briefly recall this here.

Suppose we are given a unitary, compact CFT $\mathcal{Q}$ which admits a continuous family of (compact) conformal boundary conditions $\mathcal{B} (\lambda)$ with $\lambda \in X$.
The central charge of the theory $\mathcal{Q}$ does not have to be 26.
We refer to the space $X$ as a ``boundary conformal manifold.''
Boundary conditions at different points on $X$ are connected by exactly marginal boundary perturbations.

We assume that at a generic point on $X$, the boundary condition $\mathcal{B} (\lambda)$ is simple and does not decompose into a direct sum of other boundary conditions.\footnote{There are examples where $X$ contains singular loci along which the conformal boundary condition becomes non-simple \cite{Gaberdiel:2001zq,Choi:2025ebk}.}
It follows then, given two generic points $\lambda_1 , \lambda_2 \in X$, there is a unique bcc operator between $\mathcal{B}(\lambda_1)$ and $\mathcal{B}(\lambda_2)$ with the lowest scaling dimension if $\lambda_1$ and $\lambda_2$ are sufficiently closed to each other.
We denote such a bcc operator by $\psi_{\lambda_1 \lambda_2}$ and its scaling dimension by $\Delta_{\lambda_1 \lambda_2}$.
Given three nearby points $\lambda_1 , \lambda_2 , \lambda_3 \in X$, we consider the 3-point correlation function of bcc operators on a unit disk,
\begin{align}
\begin{split}
    &\langle \psi_{\lambda_1 \lambda_2 } (\theta_1 ) \psi_{\lambda_2 \lambda_3 } (\theta_2 ) \psi_{\lambda_3 \lambda_1 } (\theta_3 ) \rangle_{D^2} \\
    &~~ = c(\lambda_1 , \lambda_2 , \lambda_3 ) \left[
        2 \mathrm{sin} \left(
            \frac{\theta_2 - \theta_1}{2}
        \right)
    \right]^{-\Delta_{\lambda_1 \lambda_2 \lambda_3}}  \left[
        2 \mathrm{sin} \left(
            \frac{\theta_3 - \theta_2}{2}
        \right)
    \right]^{-\Delta_{\lambda_2 \lambda_3 \lambda_1}} \left[
        2 \mathrm{sin} \left(
            \frac{\theta_1 - \theta_3}{2}
        \right)
    \right]^{-\Delta_{\lambda_3 \lambda_1 \lambda_2}} \,,
\end{split}
\end{align}
where $c(\lambda_1 , \lambda_2 , \lambda_3 )$ is the OPE coefficient of the three bcc operators, and $\Delta_{\lambda_1 \lambda_2 \lambda_3} = \Delta_{\lambda_1 \lambda_2} +\Delta_{\lambda_2 \lambda_3} -\Delta_{\lambda_3 \lambda_1}$, and so on.
We write
\begin{equation}
    c(\lambda_1 , \lambda_2 , \lambda_3 ) = |c(\lambda_1 , \lambda_2 , \lambda_3 )|e^{i\phi (\lambda_1 , \lambda_2 , \lambda_3 )} \,, \quad \phi (\lambda_1 , \lambda_2 , \lambda_3 ) \in \mathbb{R} / 2\pi \mathbb{Z} \,.
\end{equation}

The proposal of \cite{Choi:2025ebk} was to define a 2-form connection on $X$ by
\begin{equation} \label{eq:tilde_B}
    \widetilde{B}_{ij} (\lambda )= -\mathrm{i} \left[
        \frac{\partial^2 \phi (\lambda_1 , \lambda_2 , \lambda_3 )  }{\partial \lambda_2^{i} \partial \lambda_3^{j} } - (i \leftrightarrow j)
    \right]_{\lambda_1 = \lambda_2 = \lambda_3 = \lambda} \,. 
\end{equation}
If we use the theory $\mathcal{Q}$ as a part of the worldsheet CFT, then the family of conformal boundary conditions $\mathcal{B} (\lambda)$ gives rise to the corresponding family of on-shell open string field configurations $\Psi (\lambda ; \lambda_0)$ with respect to some chosen reference point $\lambda_0 \in X$ around which we formulate the open string field theory.\footnote{If the worldsheet matter CFT composes of $\mathcal{Q}$ with central charge $c_{\mathcal{Q}}$ and another CFT $\mathcal{Q}'$ with central charge $c_{\mathcal{Q}'} = 26 - c_{\mathcal{Q}}$, we also need to fix a reference conformal boundary condition of $\mathcal{Q}'$. We do not change this reference boundary condition of $\mathcal{Q}'$ and only vary the boundary condition of $\mathcal{Q}$, and obtain a family of on-shell open string fields $\Psi (\lambda;\lambda_0)$.}
It is plausible that the 2-form connection $B_{ij} (\lambda)$ defined in this work using the non-commutative algebra of open string field theory and the above $\widetilde{B}_{ij} (\lambda)$ are equivalent in this case.
This might be challenging to demonstrate, since in general even if $B_{ij}$ and $\widetilde{B}_{ij}$ were equivalent, the relation between them may involve a nontrivial coordinate transformation of $\lambda$ (modulo gauge transformation and also an overall rescaling of the normalization of the 2-form connections).
Results from \cite{Kudrna:2012um} can potentially be useful for a detailed comparison between $B_{ij}$ and $\widetilde{B}_{ij}$.

\section{Sigma models on boundary conformal manifolds} \label{sec:sigma_models}

One of our motivations was to understand different ways in which correlation functions on the boundary are somehow enough to reconstruct the full bulk two-dimensional CFT.
In this Section, we briefly discuss simple examples where this is possible in an interesting way.
The 2-form connection $\widetilde{B}_{ij}$ in \eqref{eq:tilde_B}, defined in the space of conformal boundary conditions, plays a crucial role.
As in Section \ref{sec:another_version}, we consider a unitary, compact, and also bosonic CFT $\mathcal{Q}$ with a nontrivial family of conformal boundary conditions $\mathcal{B} (\lambda)$ forming a boundary conformal manifold $X$.
We also allow $X$ to be a submanifold which is embedded inside a larger conformal manifold.

Let $V_i$, $i=1, \cdots , N$, be the exactly marginal boundary operators on the conformal boundary condition $\mathcal{B} (\lambda = 0)$.
A well-known fact is that the space $X$ admits a Riemannian metric given by the 2-point function of marignal operators \cite{Zamolodchikov:1986gt,Seiberg:1988pf,Kutasov:1988xb,Drukker:2022pxk,Herzog:2023dop},
\begin{equation} \label{eq:metric}
    \langle V_i (x) V_j (0) \rangle^{\lambda} = \frac{1}{2\pi}\frac{G_{ij} (\lambda)}{ x^2} \,.
\end{equation}
The superscript $\lambda$ on the left-hand side is to indicate that we compute the correlation function in the presence of the marginal perturbation on the boundary,
\begin{equation} \label{eq:boundary_pert}
    \frac{1}{\pi^{1/2}} \sum_{i=1}^N \lambda^i \int \mathrm{d}x \, V_i (x) \,. 
\end{equation}

Given a pair $(\mathcal{Q} , X)$ of a CFT and a boundary conformal manifold, it seems natural to consider a two-dimensional non-linear sigma model with the target space $X$, whose (Euclidean) action is given by
\begin{equation}
    S= \frac{1}{4\pi} \int \left( G_{ij} (\lambda) \mathrm{d}\lambda^i \wedge \star \mathrm{d}\lambda^j - i\widetilde{B}_{ij} (\lambda) \mathrm{d}\lambda^i \wedge \mathrm{d} \lambda^j \right) \,,
\end{equation}
where now $\lambda^i$'s are dynamical fields and $\star$ is the usual Hodge star.
Let us denote such a sigma model by $\mathcal{T} (\mathcal{Q} , X)$.\footnote{Sigma models whose target spaces are usual conformal manifolds of bulk marginal deformations appear in \cite{Gomis:2015yaa}.}

Given a CFT $\mathcal{Q}$, we would like to ask if there is a choice of $X$ such that
\begin{equation} \label{eq:question}
    \mathcal{T}(\mathcal{Q} , X) \cong \mathcal{Q} \,.
\end{equation}
Here, we use $\cong$ to indicate an exact equivalence of the theories on the two sides. 
Since the sigma model $\mathcal{T}(\mathcal{Q} , X)$ is constructed purely out of the information about conformal boundary conditions and boundary correlation functions of $\mathcal{Q}$, if we can find an $X$ for which \eqref{eq:question} holds, then we would conclude that the theory $\mathcal{Q}$ is fully determined by its boundary correlation functions.

It is obvious that \eqref{eq:question} cannot be true in general.
For instance, take $\mathcal{Q}$ to be a Virasoro minimal model.
In that case, there are only finitely many conformal boundary conditions \cite{Cardy:1989ir,Cardy:2004hm}, and only possible choices for $X$ will be discrete sets of points.
It is then impossible to achieve \eqref{eq:question}.
One has perhaps a better chance of success if the original CFT $\mathcal{Q}$ itself was a sigma model.
We now discuss examples where this is the case and find an $X$ satisfying \eqref{eq:question}.
In fact, usually there are multiple possible choices for $X$.

As a warm-up, let us discuss examples where $X$ is one-dimensional, and we do not have any nontrivial $\widetilde{B}_{ij}$.
Take the CFT $\mathcal{Q}$ to be the $c=1$ free compact boson at an arbitrary radius $R$, with the action given by
\begin{equation}
    S = \frac{R^2}{4\pi} \int \mathrm{d}\phi \wedge \star \mathrm{d}\phi \,.
\end{equation}
The boson field is normalized such that $\phi \sim \phi + 2\pi$, and our convention for the radius $R$ is such that the $T$-duality acts by $R \leftrightarrow 1/R$.
A possible choice for $X$ is the space of Dirichlet boundary conditions, which is a circle.
We denote it by $S^1_{D}$.
The exactly marginal boundary operator in this case is the normal derivative of $\phi$ on the boundary, $V = \frac{R^2}{2\pi^{1/2}}\partial_{\perp} \phi$.
The overall normalization of the operator is chosen so that the marginal coupling $\lambda$, in our convention \eqref{eq:boundary_pert}, is $2\pi$-periodic.
We find that the metric on $S^1_D$ given by \eqref{eq:metric} is
\begin{equation}
    R^2 \mathrm{d}\lambda^2 \,,
\end{equation}
which is that of a circle with radius $R$. 
We then achieve \eqref{eq:question}.

A slightly more interesting possibility is to consider the family of Neumann boundary conditions.
They also form a circle, which we denote by $S^1_N$.
The exactly marginal boundary operator is now the derivative of $\phi$ along the direction of the boundary, $V = \frac{i}{2\pi^{1/2}} \partial_{||} \phi$.
The overall normalization is again chosen such that the marginal coupling $\lambda$ is $2\pi$-periodic.
In this case, the metric \eqref{eq:metric} gives us 
\begin{equation}
    \frac{1}{R^2} \mathrm{d}\lambda^2 \,,
\end{equation}
which is that of a circle with radius $1/R$.
The sigma model $\mathcal{T} (\mathcal{Q} , X = S^1_N )$ is then again equivalent to the original compact boson theory $\mathcal{Q}$ by $T$-duality.

Examples with nonzero $\widetilde{B}_{ij}$ are provided by the Wess-Zumino-Witten (WZW) models.
Take $\mathcal{Q}$ to be the diagonal theory with respect to a current algebra $\mathfrak{g}_k$ where $\mathfrak{g}$ is a simple Lie algebra and $k \in \mathbb{Z}_{\geq 1}$ is the level.
The theory is a sigma model whose target space is the simply-connected group manifold $G$ whose Lie algebra is $\mathfrak{g}$ with the metric proportional to the Killing metric and the $B$-field given by the level $k$ Wess-Zumino term.
We focus only on the cases where $G$ is compact.
The theory admits a family of conformal boundary conditions which forms the group manifold $G$, coming from imposing a Dirichlet boundary condition on the sigma model field.
We take $X = G$.

The exactly marginal boundary operators are given by the holomorphic current operators $J^a (z)$ restricted to the boundary ($a$ is the Lie algebra index) \cite{Recknagel:1998ih}.
One can alternatively consider anti-holomorphic current operators $\bar{J}^a (\bar{z})$ since $J^a$ and $\bar{J}^a$ are identified at the boundary.
In \cite{Choi:2025ebk}, it was shown that $\widetilde{B}_{ij}$ in this case coincides with the level $k$ Wess-Zumino term.
Therefore, in the case of WZW models we can again achieve \eqref{eq:question}.

\section{Discussion} \label{sec:discussion}

If the non-commutative star algebra of open string field theory were to be a valid replacement in string theory of Riemannian manifolds in classical general relativity, it must know about the massless closed string backgrounds at least when the geometric interpretation of the worldsheet CFT is clear.
The result presented in this work proposes how this may be possible in the case of the Kalb-Ramond field.

More technically, we have defined a 2-form connection $B_{ij}$ in the space of classical solutions of open string field theory.
The definition is reminiscent of Berry connection in quantum mechanics.
We have shown that the gauge transformation of the open string field as well as changing the reference conformal boundary condition on the worldsheet induce a usual trasnformation of $B_{ij}$ by a 1-form gauge parameter.
Therefore, the corresponding curvature and holonomies are gauge-invariant observables which do not depend on the choice of the open string background but only on the closed string background given by the bulk of the worldsheet CFT.
We have suggested that when a family of solutions arises from a moduli space of a $D$-instanton, we should identify $B_{ij}$ with the Kalb-Ramond $B$-field of the closed string background.

Let us mention a few remaining questions.
Perhaps the most important is to explicitly test the proposal to identify $B_{ij}$ with the Kalb-Ramond field in examples.
Furthermore, one should ask how the other massless closed string backgrounds, especially the metric, are encoded in the non-commutative star algebra of open string field theory.
We have also discussed a related previous work \cite{Choi:2025ebk}, but the precise connection to the current work is not well-understood.
Finally, it will be interesting to search for similar gauge-invariant quantities associated with families of solutions in open superstring field theory or closed (super)string field theory.

Another speculative idea is to utilize the 2-form connection $B_{ij}$ to detect singularities in the solution space of open string field theory.
In quantum mechanics, the Berry curvature diverges when there is a level-crossing and this is detected by computing the flux of Berry curvature across a two-dimensional surface enclosing the singular point.
In the case of open string field theory, a nonzero flux of the 3-form curvature $H_{ijk}$ across a closed 3-dimensional manifold may indicate the presence of a singularity in the space of classical solutions.

\section*{Acknowledgments}

I am very grateful to Juan Maldacena, Kantaro Ohmori, Leonardo Rastelli, Shinsei Ryu, Edward Witten, and especially Yifan Wang for valuable discussions and comments. I also thank Atakan Hilmi Firat for helpful correspondences. This work is supported by NSF Grant PHY-2514611 and the Fund for Memberships in Natural Sciences at the Institute for Advanced Study.

\bibliographystyle{JHEP}
\bibliography{ref}

\end{document}